\documentclass[useAMS,usenatbib,rotating]{mn2e}

\usepackage{amssymb}
\usepackage{amsmath}
\usepackage{amsfonts}
\usepackage{xspace}
\usepackage{longtable}
\usepackage{rotating}
\usepackage{lscape}

\newcommand{\HI}{\mbox{\sc H{i}}}

\newcommand{\msun}{\mbox{$M_\odot$}}

\newcommand{\kms}{\mbox{km s$^{-1}$}}

\newcommand{\tophat}{{\sc TopHat}\xspace}

\title{The Northern HIPASS catalogue - Data presentation, completeness
and reliability measures.}
\author[O. I. Wong et al.]
       {O. I. Wong,$^{1,2}$ E. V. Ryan-Weber,$^{1,3}$
	 D. A. Garcia-Appadoo,$^{4,5}$ R. L. Webster,$^1$\newauthor L. Staveley-Smith,$^2$
	 M. A. Zwaan,$^6$  M. J. Meyer,$^7$  D. G. Barnes,$^1$     V. A. Kilborn,$^{2,8}$ 
	  \newauthor R. Bhathal,$^9$ W. J. G. de Blok,$^{10}$
	 M. J. Disney,$^4$ M. T. Doyle,$^{11}$ M. J. Drinkwater,$^{11}$\newauthor R. D. Ekers,$^2$
	K. C. Freeman,$^{10}$ B. K. Gibson,$^{12}$ S. Gurovich,$^{10}$
	J. Harnett,$^{13}$ \newauthor
	P. A. Henning,$^{14}$  H. Jerjen,$^{10}$
	M. J. Kesteven,$^2$  P. M. Knezek,$^{15}$  B. S. Koribalski,$^2$
	\newauthor S. Mader,$^2$ M. Marquarding,$^2$ R. F. Minchin,$^{16}$
	J. O'Brien,$^{10}$  M. E. Putman,$^{17}$ 
	\newauthor S. D. Ryder,$^{18}$ E. M. Sadler,$^{19}$ J. Stevens,$^{20,1}$
	I. M. Stewart,$^{21}$  F. Stootman,$^9$ \newauthor and M. Waugh$^{1,2}$\\
        $^1$School of Physics, University of Melbourne, VIC 3010, Australia\\
	$^2$Australia Telescope National Facility, CSIRO, PO Box 76, 
	Epping, NSW 1710, Australia\\ 
        $^3$Institute of Astronomy, Madingley Road, Cambridge CB3 0HA,
	U.K.\\
	$^4$Department of Physics \& Astronomy, University of Wales,
	Cardiff, P.O. Box 913, Cardiff CF2 3YB, U.K.\\
	$^5$RAIUB, Radioastronomisches Institute Universitat Bonn, Auf
	dem Hugel 71, 53121, Bonn, Germany\\
	$^6$European Southern Observatory, Karl-Schwarzschild-Str. 2,
	85748 Garching b. Muenchen, Germany\\
	$^7$Space Telescope Science Institute, 3700 San Martin Drive,
	Baltimore, MD 21218, U.S.A.\\
	$^8$Centre for Astrophysics \& Supercomputing, Swinburne
	University of Technology, P.O. Box 218, Hawthorn, VIC 3122, Australia\\	
	$^9$Department of Physics, University of Western Sydney
	Macarthur, P.O. Box 555, Campbelltown, NSW 2560, Australia\\
	$^{10}$Research School of Astronomy \& Astrophysics, Mount
	Stromlo Observatory, Cotter Road, Weston, ACT 2611,
	Australia\\
	$^{11}$Deparment of Physics, University of Queensland, QLD 4072, Australia\\
  $^{12}$University of Central Lancashire, Centre for Astrophysics, Preston, PR1 2HE, U.K.\\
	$^{13}$University of Technology Sydney, Broadway, NSW 2007,
	Australia\\
	$^{14}$Institute for Astrophysics, University of New Mexico,
	800 Yale Blvd, NE, Albuquerque, New Mexico 87131, U.S.A.\\
	$^{15}$WIYN, Inc. 950 North Cherry Avenue, Tucson, Arizona,
	U.S.A.\\
  $^{16}$Arecibo Observatory, HC03 Box 54995, Arecibo 00612, Puerto Rico\\
	$^{17}$Department of Astronomy, University of Michigan, Ann Arbor, MI 48109, U.S.A\\
	$^{18}$Anglo-Australian Observatory, P.O. Box 296, Epping, NSW
	1710, Australia\\
	$^{19}$School of Physics, University of Sydney, NSW 2006, Australia\\
  $^{20}$Department of Physics, University of Tasmania, TAS 7005, Australia\\
	$^{21}$Department of Physics \& Astronomy, University of
	Leicester, Leicester LE1 7RH, U.K.\\
}

\begin{document}

\date{Accepted ***. Received ***; in original form ***}

\pagerange{\pageref{firstpage}--\pageref{lastpage}} \pubyear{2003}

\maketitle
 
\label{firstpage}

\begin{abstract}

The Northern HIPASS catalogue (NHICAT) is the northern extension
of the HIPASS catalogue, HICAT \citep{meyer04}.  This  extension
 adds the sky area between the declination range of $+2^{\circ} < \delta < +25^{\circ}30\arcmin$
to HICAT's declination range of $-90^{\circ} < \delta < +2^{\circ}$.  HIPASS is a blind
\HI\ survey using the Parkes Radio Telescope covering $71\%$ of the sky (including
this northern extension) and a heliocentric velocity range of -1,280 \kms\ to 
12,700 \kms\ . The entire Virgo Cluster region has been observed in the Northern 
HIPASS. The galaxy catalogue, NHICAT, contains {\em{1002}} sources  with 
$v_{\rm{hel}} > 300$ \kms\ .   Sources with $-300$ \kms\ $< v_{\rm{hel}} < 300$ \kms\ 
were excluded to avoid contamination by Galactic emission.  In total, the entire HIPASS survey has found 
{\em{5317}} galaxies  identified purely by their HI content. The full galaxy catalogue is 
publicly-available  at  $\langle${\bf{\tt http://hipass.aus-vo.org}}$\rangle$.

\end{abstract}


\section{Introduction}

The \HI\ Parkes All-Sky Survey (HIPASS) survey is a blind \HI\ survey
using the Parkes Radio Telescope\footnote{The Parkes telescope is part of the Australia
  Telescope  which is funded by the Commonwealth of
  Australia for operation as a National Facility managed by CSIRO.}, 
and the Northern extension increases this survey by a further 37 percent in sky area. 
The primary objective of extending  Southern HIPASS to the north is to complement
the southern census of gas-rich galaxies in the local Universe.

The \HI\ mass function, HIMF, \citep{zwaan03} and galaxy two-point
correlation function \citep{meyer05} based on Southern HIPASS
showed that the statistical measures of
the galaxy population from HIPASS are limited by cosmic variance.  
Recently, \citet{zwaan05} used HICAT to investigate the effects
of the local galaxy density on the HIMF.  Using the $n$-th nearest neighbour
statistic, they
found  tentative evidence that the low-mass end of the HIMF becomes steeper in higher
density regions.   These authors were able to
examine the trend in the slope of the HIMF for different values of $n$ in
the statistic.  Larger values of $n$ correspond to sampling the density
on larger scales.  For each value up to $n=10$, the slope became systematically steeper
as the density increased.
Thus it appears that the \HI\
properties of galaxies might be influenced by environmental effects on quite large
scales (where a typical separation of the $n=10$ nearest neighbours is $\sim$ 5 Mpc), in addition to the well-known local effects, such as tidal interactions
between neighbouring galaxies.
Previously, \citet{rosenberg02}  found $\alpha \approx -1.2$ and $\alpha \approx -1.5$ for the slope of the HIMF in the immediate and field regions of the Virgo Cluster, respectively.  Also, \citet{spring05}, found flatter slopes to the low mass end of the HIMF
in higher density regions.  However their galaxy sample was selected optically.  Since Northern HIPASS covers the entire Virgo
Cluster region,  the Northern HIPASS catalogue (NHICAT) can be used in conjunction
with the Southern HIPASS catalogue (HICAT) to explore these trends and investigate the
effects of cosmic variance on HIPASS galaxy catalogue statistics.

It is worth noting that the Northern HIPASS also provides 
the first blind \HI\ survey of the entire region in and
around the Virgo Cluster.   Assuming a Virgo distance of 16 Mpc and
 an integrated flux limit of 15 Jy \kms\, 
this corresponds to a mass sensitivity of $9 \times 10^8$ \msun.  
Thus the survey will detect any \HI\ clouds above this mass limit in the
vicinity of the Virgo Cluster, regardless of stellar content.  

The Virgo Cluster provides a nearby example of processes
that are more common at higher redshifts, such as galaxy-galaxy and
galaxy-intracluster medium interactions. Northern HIPASS will be used
to investigate the role of \HI\ in a cluster environment in individual
galaxies as well as statistically across the whole
cluster. Understanding the role of \HI\ is vital for galaxy evolution
models.     \citet{kenney04} found
6 galaxies in the Virgo Cluster showing distorted
\HI\ morphology.  Using N-body simulations, \citet{vollmer01} investigated the 
effect of ram pressure stripping in the Virgo Cluster and found that
\HI\ deficiency is dependent on galaxy orbits within the cluster.   They 
concluded that all the galaxies showing some form of distorted \HI\
distribution have already passed through the centre of the cluster and
are not infalling for the first time. 

The catalogue of extragalactic \HI\ sources from HIPASS was named HICAT
and was presented in \citet{meyer04} (hereafter known as MZ04), while the completeness and
reliability of HICAT was assessed by \citet{zwaan04}. Here we present a
catalogue of extragalactic \HI\ sources from Northern 
HIPASS, named NHICAT. The basic parameters of HIPASS, Northern
HIPASS, HICAT and NHICAT are given in Table~\ref{params}. Apart from the
declination coverage, the main difference between the two surveys and
catalogues is the higher noise level in Northern 
HIPASS.  For a full summary of parameters of
existing blind \HI\ surveys, including subsets of HIPASS (HIPASS Bright
Galaxy Catalogue, \citet{korib04}; the South Celestial Cap catalogue,
\citet{kilborn02}) see Table 1 of MZ04.  The Northern
HIPASS Zone of Avoidance (NHIZOA) survey by \citet{donley05} covers 
northern declinations of the Galaxy - a subset of the Northern HIPASS area - at
a higher sensitivity  (RMS = 6 mJy beam$^{-1}$).

Optical identification of NHICAT sources will use similar
techniques to HICAT \citep{doyle05} and will be presented in a later paper.
With a total 
spatial coverage of 29,343 square degrees and 5317 sources, the combined
HICAT and NHICAT catalogue is the largest purely \HI-selected galaxy
catalogue to date.  The  Arecibo L-Band Feed Array (ALFALFA) surveys 
will eventually cover the same region of sky as Northern HIPASS and will 
extend up to a declination of $+36^{\circ}$. More information about the 
progress of ALFALFA can be found online at  
$\langle${\bf{\tt http://egg.astro.cornell.edu/alfalfa}}$\rangle$ \citep{giov05}.


In this paper we present NHICAT, together with the completeness
and reliability analysis of the catalogue.  Section 2 reviews 
Northern HIPASS and its properties.  The source identification and the generation 
of the catalogue is described in  Section~\ref{datasection}.  Section~\ref{noisesection}
discusses the noise statistics of Northern HIPASS and the
completeness of NHICAT is analysed in Section 3.1.  The narrowband
follow-up observations and the reliability of NHICAT will be 
described in Section 3.2.

\begin{table*}
\begin{center}
\caption{Survey and catalogue parameters.}
\label{params}
\begin{tabular}{lccccc}
\hline
Survey name & Survey range & RMS & Catalogue Name & Catalogue range 
& Sources \\
& ($deg$, \kms\ ) & (mJy beam$^{-1}$) & & ($deg$, \kms\ ) & \\
\hline
HIPASS & $\delta < +2^{\circ}$, & 13 & HICAT & $\delta < +2^{\circ}$, & 4315\\
(MZ04) & $-1280<v<12 700$ &  & $300<v<12 700$ & &\\
\\
Northern extension& $+2^{\circ} < \delta < +26^{\circ}$, & 14 & NHICAT &
$+2^{\circ} < \delta < +25^{\circ}30\arcmin$, & 1002$^{*}$\\
HIPASS, this work & $-1280<v<12 700$ & & $300<v<12 700$ &&\\
\\
\hline
\end{tabular}
\end{center}
\begin{flushleft}
$^{*}$  Note that 1 source was found slightly below declination $+2^{\circ}$. 
\end{flushleft}
\end{table*}

\section{Northern HIPASS Data}

Northern HIPASS was designed to cover all RAs in the region between declinations
$+2^{\circ} < \delta < +25^{\circ}$.   Observations were undertaken using
the 21-cm Multibeam receiver \citep{staveley96} on the Parkes radio
telescope during the period from 2000 to 2002.  Observations were made
by scanning in $8^{\circ}$ strips of sky with 7 arcminutes in RA of
separation between scans.  A 1024 channel configuration
covering a 64 MHz bandwidth was used in the Multibeam
correlator to give a channel separation of $\Delta v$ = 13.2
\kms\ across the heliocentric velocity range of -1,280 \kms\ to  12,700 \kms\ .
The observation and reduction methods  are exactly the same as
Southern HIPASS and can be found in detail in \citet{barnes01}. The northern dataset
consists of 102 $8^{\circ}\times8^{\circ}$ cubes and 48
$8^{\circ}\times7^{\circ}$ cubes in the northernmost declination band. 

The catalogue includes sources in the range $+25^{\circ}<\delta <+25^{\circ}30\arcmin $.
At this declination, the telescope field of view is increased, though the sensitivity is significantly decreased.


\subsection{Northern HIPASS Catalogue (NHICAT)}
\label{datasection}

NHICAT was generated using essentially the same method as HICAT, 
with some improvements in efficiency.  An updated version of the \tophat
finder algorithm (see MZ04 for details of the original \tophat) was used to
identify sources.  The updated \tophat finder is very effective at filtering 
false detections which have narrow velocity widths.  Velocity widths with a FWHM
of less than 30 \kms\ were considered to be too narrow to be extragalactic sources. 
Such narrow velocity width detections
are usually associated with hydrogen recombination frequencies or known 
interference lines.  The consistency of the updated finder was tested by comparing 
the output of the original and updated finder for the southern cubes.  The 
updated finder returned exactly the same sources as the original version without 
the narrow velocity width detections.  Two galaxy finders were used to generate 
HICAT: `MULTIFIND' and \tophat.  The `MULTIFIND' finder uses a peak flux
threshold method, whereas the \tophat algorithm involves the cross-correlation of 
spectra with tophat profiles at various scales (MZ04).  Although the original version of \tophat
was reported to find $\sim$90$\%$ of final catalogue sources in southern HICAT
(MZ04), further tests showed that all sources with $S_{\rm{peak}}$ $>$ 100 mJy were 
recovered by the updated \tophat finder.  Since `MULTIFIND' generated many more 
false detections in the northern data due to the different level and character of noise (see 
section 2.2), we decided to use the updated \tophat finder only.  The \tophat finder 
was found to be quite robust against the increased level of noise and baseline
ripple.  In addition, since the updated \tophat finder is much more effective,
separate radio-frequency intereference (RFI) and recombination line removal (as described in MZ04) was not
necessary.

Although the known narrowband RFI have been filtered out in the process
described above, not all the known RFI have been removed from the data.  Not only is the 
 Sun (and the reflections of the Sun) the strongest source of interference but it also
provides broadband interference which could not be easily filtered out.  This solar
interference produces a standing wave effect (known as `solar ripple' ) in our data which
in turn will affect the effectiveness of galaxy detection.  These standing waves are likely to
be worse at lower elevations due to the ground reflection of the Sun.

\begin{figure*}
\begin{center}
\vspace{14pc}
\begin{tabular}{cccc}
 & & &\includegraphics{a.ps}\\
 & &\includegraphics{b.ps} &\includegraphics{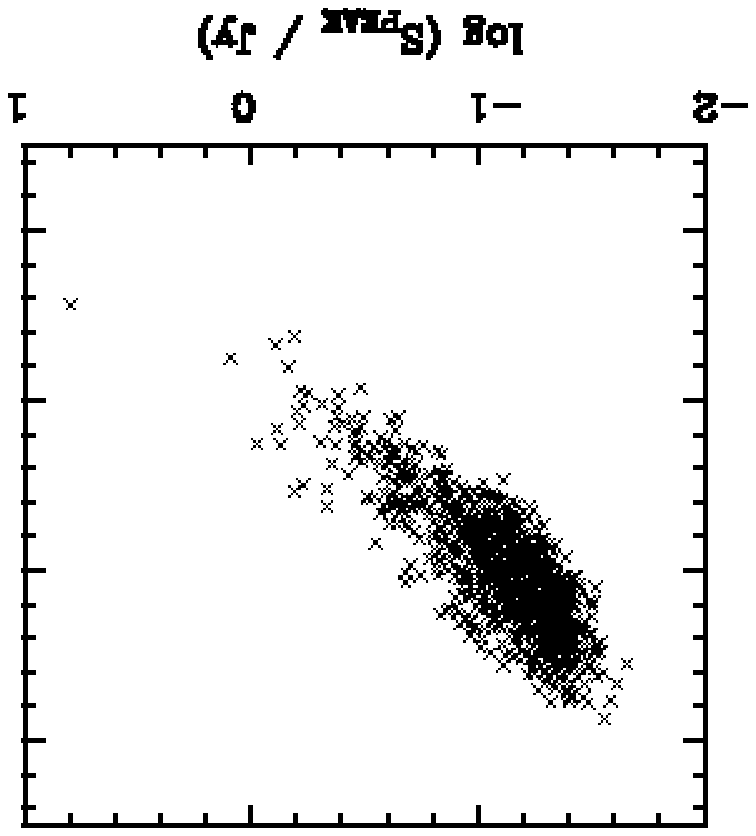} \\
 &\includegraphics{d.ps} &\includegraphics{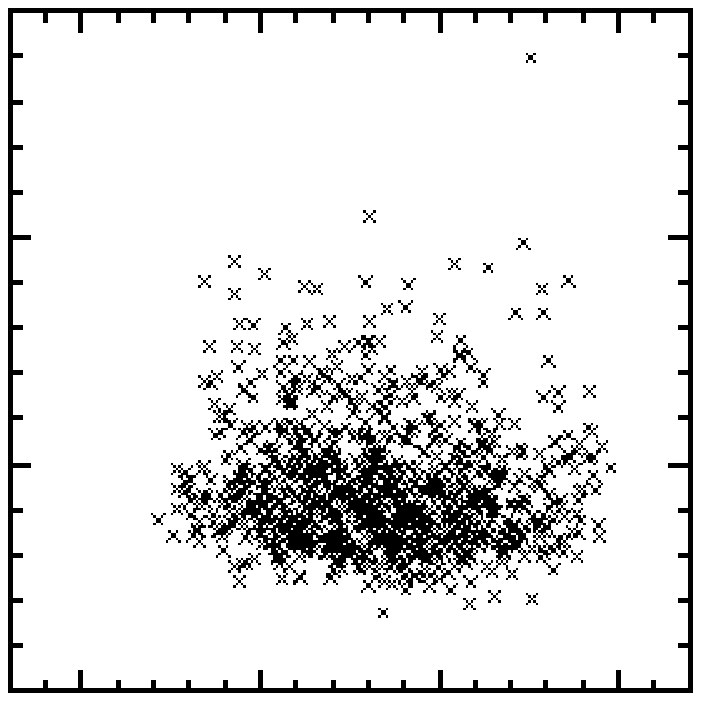} &\includegraphics{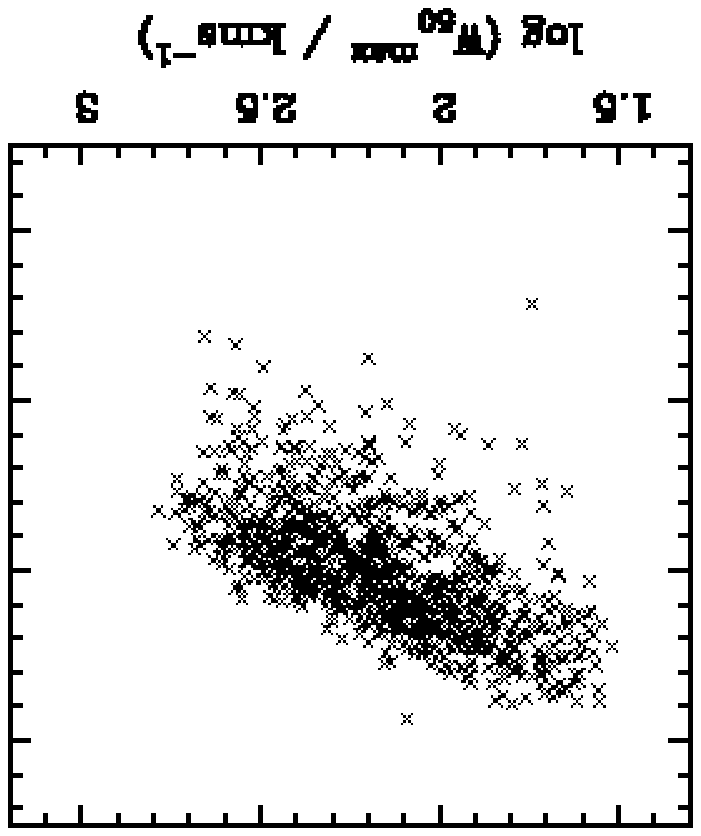}\\
\includegraphics{g.ps}& \includegraphics{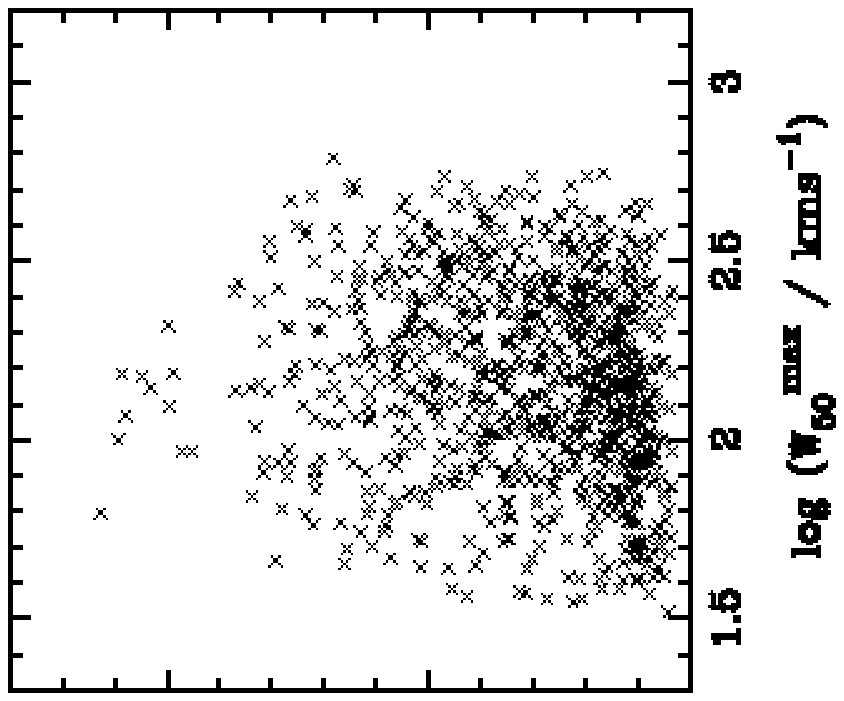} & \includegraphics{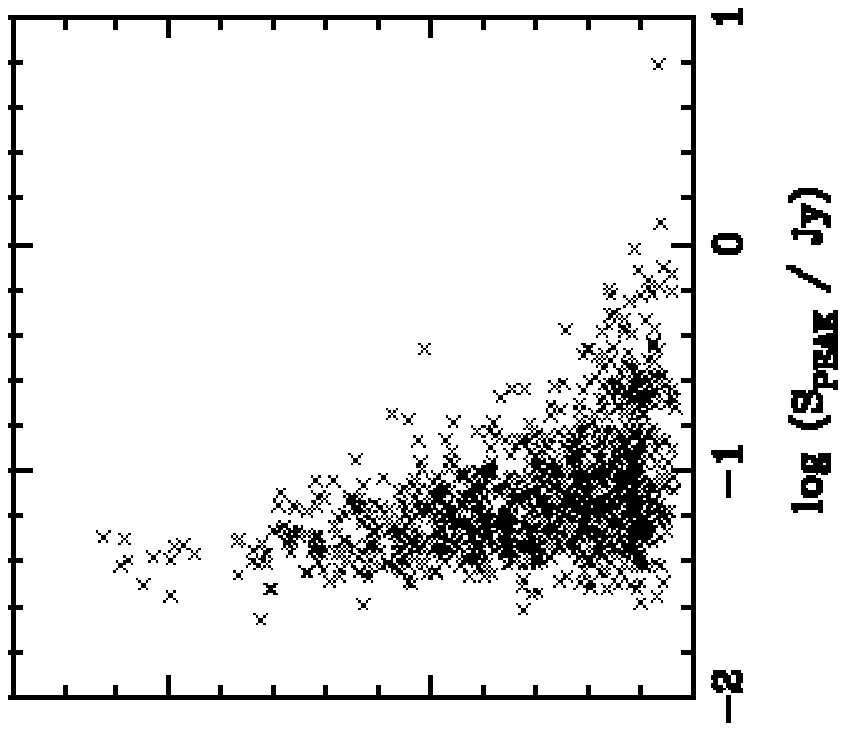} & \includegraphics{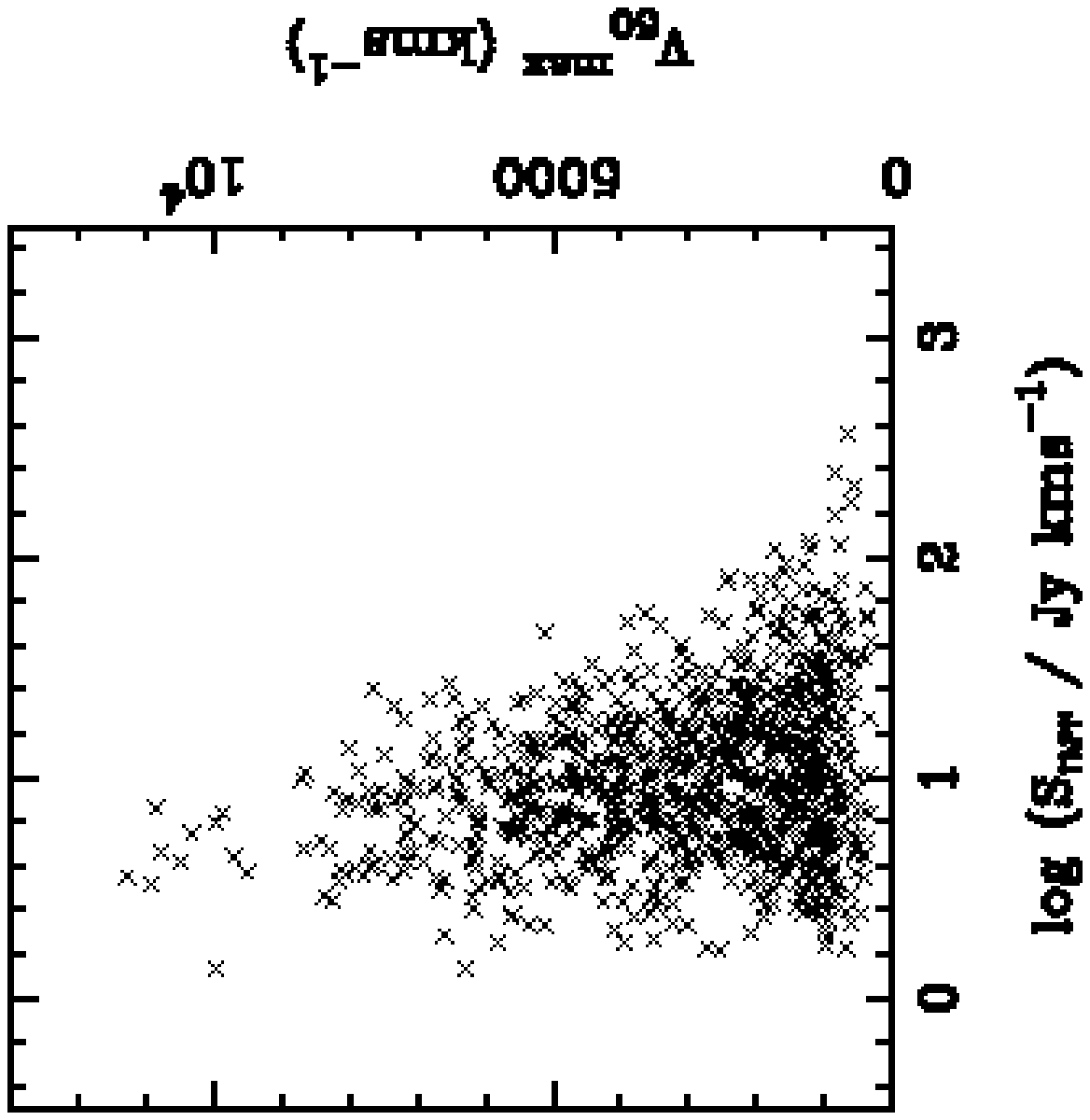} \\
\end{tabular}
\vspace{23pc}
\caption{Log-log bivariate distributions of   velocity width ($W_{\rm{50}}^{\rm{max}}$), peak flux ($S_{\rm{p}}$) and integrated
  flux ($S_{\rm{int}}$).  Plotted along the diagonal are single parameter histograms.}
\label{paramdistrib}
\end{center}
\end{figure*}

NHICAT was constructed by first running the \tophat\ galaxy finder on
all cubes, excluding the region $-300 < v_{\rm{hel}} < 300$ \kms\ to minimise
confusion with Galactic emission.  14,879 galaxy candidates were found. Each candidate source was then
checked manually by simultaneously displaying the source in four ways: 
a spectrum, an integrated intensity map, a declination-velocity projection and a 
Right Ascension-velocity projection. A candidate source was accepted if it had a spectral
profile which was easily distinguishable from the noisy baseline and a
position-velocity profile which was wider than 2 pixels across the
position axes.  On the other hand, a candidate source was rejected if either
its spectral profile was not distinguishable from the noise or if its
position-velocity projections showed a distinct signal which was
exactly 2 pixels wide, as these sources
are  usually the result of 
interference in the data.   All the accepted sources
 were then passed to a semi-interactive parameter finder.  
Sources were flagged during parameter finding as either 1, 2, 3, 4 or
 a high-velocity cloud (HVC) detection: Flag=1 represented a definite source detection, 2 represented
 source detection with less certainty, 3 represented a source detected
 on the edge of a cube and 4 represented a non-detection.  The Flag=4
option is provided to the parameter finder in the case of misclassification
of a source during the checking stage. It should be
 noted that a significant number of HVCs were detected in Northern
 HIPASS at RA $\sim$ 23 hours due to the Magellanic stream.  The Magellanic
Stream from the Northern HIPASS data has been presented by \citet{putman03}.
Source lists from all cubes were then merged and duplicates were removed.  The
 process of merging matched every Flag=3 source with the detection
 of the same source in the neighbouring cube with the same overlap
 region.  The procedures  for the manual checking and parametrisation used
are exactly the same as MZ04. More detailed descriptions of these
procedures can be found in sections 3.2 and 3.3 of MZ04.
Bivariate distributions of the basic properties (heliocentric 
velocity, velocity width, peak
flux and integrated flux) of the sources detected in
Northern  HIPASS are shown in Figure~\ref{paramdistrib}.
Single parameter histograms are shown on the diagonal of this Figure.

The cubes were then re-checked  for  missed sources
using a semi-interactive process.  Sources
detected in each
cube were marked with a cross on an integrated intensity map.  These maps were then manually checked 
for unmarked sources.  From these manual checks, 15 additional sources were
detected.  The galaxy finder missed 
sources which were close to several other sources.  

Two sources may also have been identified as the same source.  Such
sources can be differentiated by inspecting the spectra and the
spatial moment maps since separate sources will not overlap
both spatially and spectrally.  These particular sources were flagged
as `confused'.  There are two `confused' sources in NHICAT.

Extended sources were identified and fitted in the same fashion
as in MZ04 (see section 3.5 of MZ04 for more details).  All sources
greater than $7\arcmin$ in size were identified as potentially extended
sources.  The integrated flux limit corresponding to a fixed source size
can be determined using the relationship  between integrated flux
$S_{\rm{int}}$, and source diameter,

\begin{equation}
S_{\rm{int}}=\int S dV \approx 1.2 \theta^{2}_{\HI}
\label{eq1}
\end{equation}
An explanation and derivation of Equation~\ref{eq1} can be found in MZ04. 
The $S_{\rm{int}}$ limit corresponding to a source size greater than $7\arcmin$ is
57 Jy \kms\ .  In total, 41 candidates were found to have a
measured  flux greater than
this limit.  The moment maps of each of these sources were then
examined and two sources were found to be extended.

All the cube velocities are heliocentric and use the optical definition where velocity,$v=c(f_o/f)/f$ where $c$, $f_o$ and $f$ represent the speed of light, rest and observed frequency respectively. The sources were then assigned names according to the convention of HICAT \citep{meyer04}.

The final stage of the processing involved checking the 19 sources located 
at $\delta < +2^{\circ}$.   Of these, 18 sources have already been catalogued
in HICAT showing that the datasets and analysis techniques are consistent  between the two surveys at the
95\% level.  The extra source is located only slightly
south of  $\delta =+2^{\circ}$ and  has been retained in NHICAT so that the final combined
NHICAT and HICAT catalogue will be a complete catalogue of the southern skies 
up to $\delta = +25^{\circ}30\arcmin$. 


Northern HIPASS covers 7997 square degrees of sky and 1002 galaxies
were found from their \HI\ content in this region.   The
parameters  provided with the catalogue are the same as the
parameters given in HICAT.  Detailed descriptions of each parameter 
can be found in Table 4 of MZ04.
\begin{figure}
\vspace{12pc}
\includegraphics{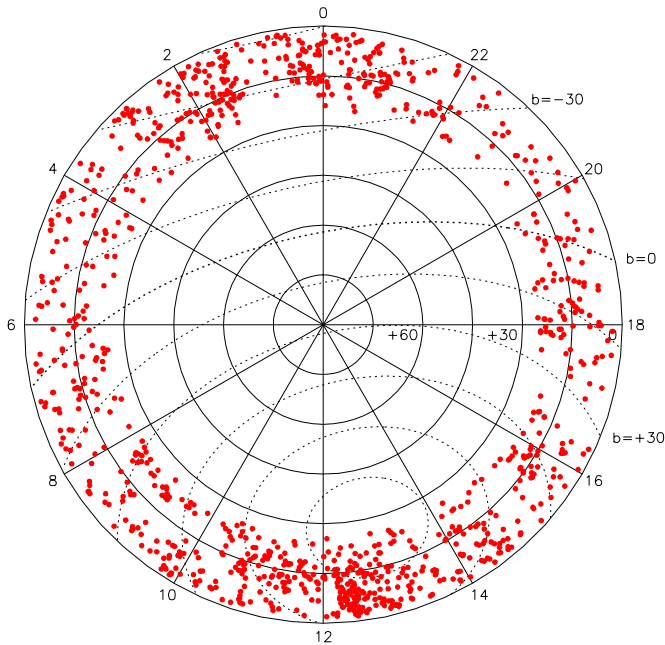}
\vspace{9pc}
\caption{Skymap of detections found in Northern
  HIPASS.  The lines mark increasing declinations inwards
  where the centre is the north pole. The radial
  divisions show increasing RA in an anti-clockwise direction
  starting with 0 hours at the top of the diagram.  The dotted
  lines mark lines of  Galactic latitude, $b$.}
\label{nhicatmap}
\end{figure}
Figure~\ref{nhicatmap} shows the spatial distribution of all the sources
found in NHICAT.  Note that the cluster of sources at RA, $\alpha \sim
12$ hours is the Virgo cluster.   Galaxies are also found at low Galactic
latitudes - a region often avoided by optical galaxy surveys.  As mentioned
in the introduction, NHIZOA (\citet{donley05}) covers a subset of the 
Northern HIPASS survey region.  We re-identify 23 galaxies from the 
NHIZOA catalogue in the overlapping regions.  
The RMS of NHIZOA is 2.33 times less than that of the observations in 
Northern HIPASS.  Assuming that the detected number of galaxies is 
based solely  on the sensitivity of the observations, one would 
expect Northern HIPASS to find only 33 of the 77 NHIZOA sources.
There are  fewer matches between Northern HIPASS and NHIZOA because
 the source detection rate is not fully described by the sensitivity (as explained 
in the following section).

In accordance with HICAT conventions, NHICAT only included 
detections with $v_{\rm{hel}} > 300$ \kms\ .  One galaxy,
HIPASSJ1213+14a, was found to have a mean \HI\ $v_{\rm{hel}}$ of -222.7
\kms\ , $S_{\rm{peak}}=0.181$ Jy and $S_{\rm{int}}=42.9$ Jy \kms\.
This galaxy was detected because its velocity width extended into the
heliocentric velocity range where  $v_{\rm{hel}} < -300$ \kms\.  In summary, no galaxies
with $v_{\rm{hel}} < -300$ \kms\ were found in NHICAT.  The number 
density of the sources found in Northern HIPASS was approximately 
0.13 sources per square degree of sky.  In comparison, 0.20 sources per 
square degree of sky were found in Southern HIPASS.  The cause of this 
difference will be discussed later in this paper.

\begin{table*}
\centering
\caption{Comparison of source number  density in NHICAT and
  HICAT. Note that 1 source in NHICAT was detected below
  $2^{\circ}$ in declination.}
\label{ndense}
\begin{tabular}{lcccc}
\hline
\hline
Catalogue & Declination range & Area of sky (sq degs) & No.\ of sources
& Number density (per sq degs) \\ 
\hline
HICAT & $-90^{\circ}< \delta < +2^{\circ}$& 21,346 & 4,315 &0.20\\
\hline
NHICAT & $+2^{\circ} \leq \delta < +10^{\circ}$&2,862 & 413 & 0.14\\
  & $+10^{\circ} \leq \delta < +18^{\circ}$&2,792 & 364& 0.13 \\
  & $+18^{\circ} \leq \delta < +25^{\circ}$&2,343 & 224 & 0.10 \\
\hline
\hline
\end{tabular}
\end{table*}


\subsection{Noise characteristics}
\label{noisesection}

Fewer sources were detected in NHICAT at higher declinations than in HICAT, as can be seen in Table~\ref{ndense} (which shows the number density in NHICAT and HICAT).  Although some deviation is expected due to cosmic variance, the most likely cause of this density difference is the higher level of noise in Northern HIPASS.  The variation in gain and system temperature ($T_{\rm{sys}}$) of the telescope with respect to elevation are insufficient to account for the higher noise observed in the northern survey.  The level of noise in Northern HIPASS is greater than in
Southern HIPASS because the Parkes radio telescope observes northern sources at
lower sky elevations. This results in the telescope gathering a
greater amount of interference from ground reflections and the Sun.  As the most 
northerly areas of the survey were only observable during a short LST window, 
sidelobe solar interference was often unavoidable.

\begin{figure}
\includegraphics{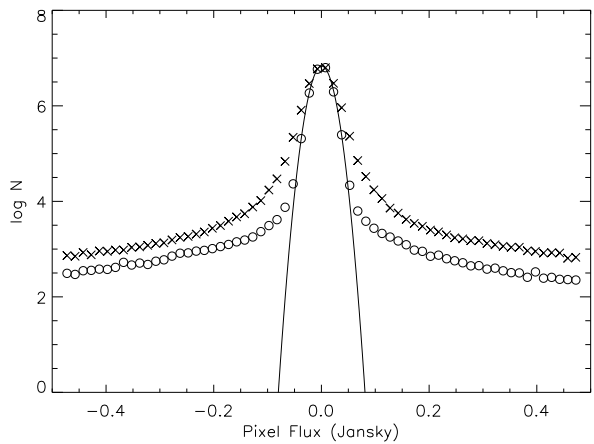}
\vspace{15pc}
\caption{Peak-normalised distributions of pixel flux of entire cubes in HIPASS.  
The distribution of the pixel
  flux from a typical northern cube (cube number 538) is represented by crosses and  the
  distribution marked by asterisks represent the pixel flux from a typical
  southern cube (cube number 194).  A parabola (solid line) is shown to compare these 
distributions with Gaussian noise statistics.}
\label{pixfluxdist}
\end{figure}

As shown in Table~\ref{params}, the RMS of both Northern and Southern
HIPASS are very similar.  However, the Northern HIPASS cubes
appeared much `noisier' in the flux density maps than that of the cubes in Southern HIPASS. Hence, the
RMS method is not an effective way of illustrating why Northern HIPASS
appeared `noisier'.
\begin{figure}
\vspace{12pc}
\includegraphics{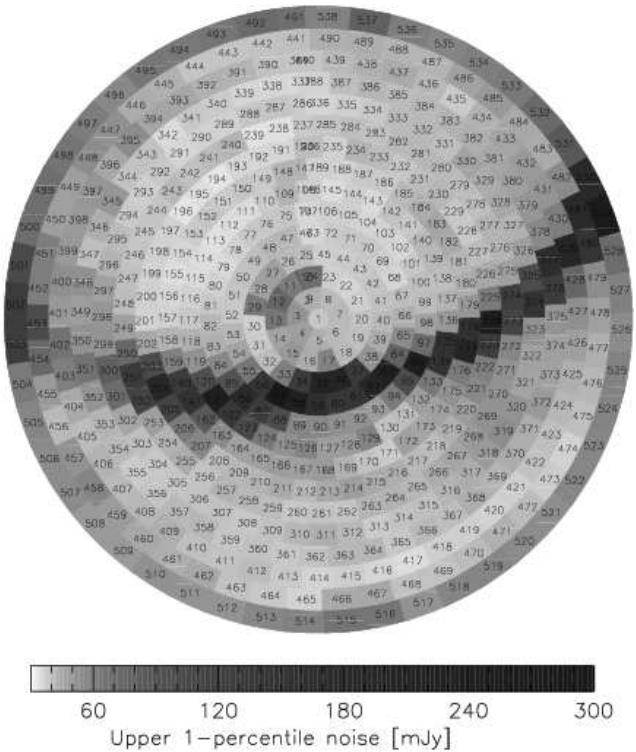}
\vspace{11pc}
\caption{Skymap of the 1-percentile pixel flux map of both Northern
and Southern HIPASS.  The south pole ($\delta=-90^{\circ}$) is in the
centre and RA increases in an anti-clockwise direction 
starting with 0 hour at the top of the diagram.  Observations through the 
Galaxy (where $b=0^{\circ}$) correspond to the darker horizontal band of cubes.  
The southern cube identification numbers range
  from 1-388 and the northern cubes are 389-538.}
\label{noisemap}
\end{figure}
 
\begin{figure}
\includegraphics{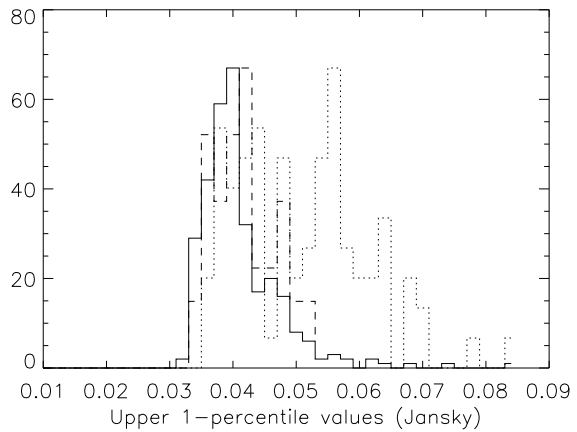}
\vspace{14pc}
\caption{Normalised histograms of `noise' levels (1-percentile measures). The distribution of `noise'
  levels in Southern HIPASS is represented by the solid line distribution.  The
  dashed line distribution represents the `noise' levels in the
  declination band between $+2^{\circ} < \delta < +10^{\circ}$ and
  the dotted line distribution represents the `noise' levels in the
  declination bands between  $+10^{\circ} < \delta < +26^{\circ}$.}
\label{noisehist}
\end{figure}
 
The aim of this section is to characterise the distribution of noise observed in
Northern HIPASS cubes.  One method is to examine the
distribution of all the pixel flux values in a given cube and
measure the 99-percentile value of this distribution.  This measure illustrates the
noise characteristics in a given cube by measuring the extent of the 
outlying pixel flux values in the pixel distribution.  Instead of characterising the 
noise in terms of the width of the flux distribution (as in the RMS method), we
now examine the extent to which the outlier population is extended.

Figure~\ref{pixfluxdist} shows the peak-normalised pixel flux distributions of 2
average cubes, one from Northern HIPASS (cube number 538) and the other from 
Southern HIPASS (cube number 194).
Since \HI\ detections comprise very few pixels compared with the
entire cube, the \HI\ sources have not been removed  from the plot.
The excess of negative flux in Northern cubes (as seen in Figure~\ref{pixfluxdist})
 is due to the bandpass removal and calibration method.  Negative bandpass
 sidelobes occur at declinations either side of bright sources \citep{barnes01}.  This
 means that bright interference is surrounded by negative artifacts, leading to an 
excess of both positive and negative pixels in the data with stronger interference.
An ideal cube with only Gaussian noise would have a parabolic
distribution (as shown by the solid line) since the natural log of a
Gaussian distribution,  exp($-x^2$), is  $-x^2$.
As can be seen in Figure~\ref{pixfluxdist}, the offset from the parabola is 
greater in the Northern cube than in the Southern cube, suggesting 
broader distributions of pixel noise values in the North.

Thus the `noise' level can now be characterised by measuring the extent of outliers
using the 99-percentile rank of the pixel flux distribution.  In actuality, the  
$1$-percentile rank has been used to
avoid bias caused by actual \HI\ sources in this measure.
Figure~\ref{noisemap} shows the skymap of  the $1$-percentile
measures for both Northern
HIPASS and Southern HIPASS.   
It should be noted that the 3 outer bands of declinations (cube
numbers 389 -
538) form the regions observed in Northern HIPASS,  and  the inner
declination bands (cube numbers 1 - 388) represent the observed regions in Southern HIPASS.

It is qualitatively obvious from Figure~\ref{noisemap} that observations through
the Galaxy appear much `noisier' than observations away from the Galaxy.  It can 
also be seen that the northernmost declination band in Northern HIPASS is much
noisier than the rest of the HIPASS observations.

\begin{figure*}
\begin{center}
\includegraphics{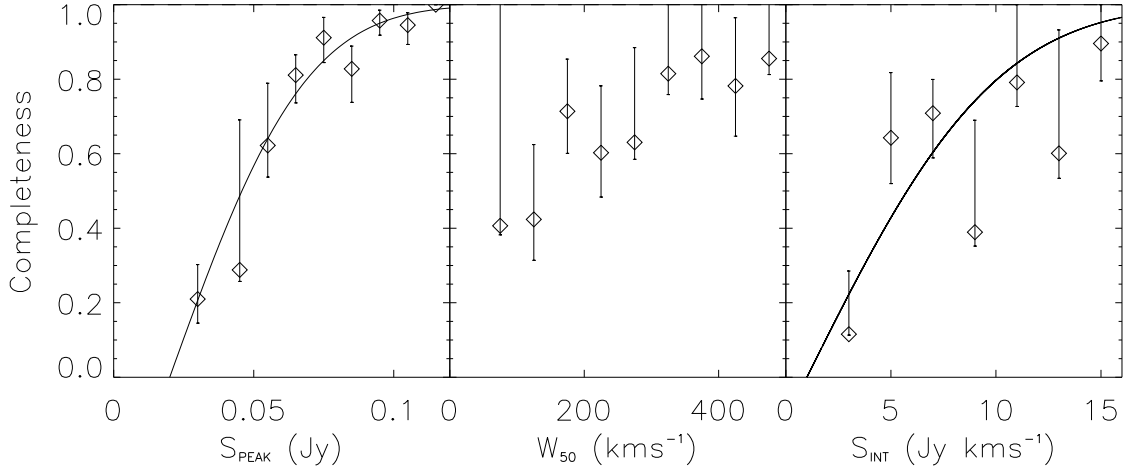}
\vspace{15pc}
\caption{Completeness of NHICAT as a function of $S_{\rm{p}}$, $W_{\rm{50}}$ and $S_{\rm{int}}$.}
\label{compl}
\end{center}
\end{figure*}

\begin{table*}
\caption{Completeness of NHICAT.}
\label{comp}
\centering
\begin{tabular}{llc}
\hline
Parameter & Completeness Fit & C=0.95 \\
\hline
$S_{\rm{p}} $ (Jy) & erf$[18.5(S_{\rm{p}}-0.02)]$ & 0.095\\
$S_{\rm{int}}$ (Jy \kms\ ) & erf$[0.1(S_{\rm{int}}-1.0)]$&15.0 \\
\hline
$S_{\rm{p}} $(Jy), $S_{\rm{int}}$ (Jy \kms\ ) &erf$[20.0(S_{\rm{p}}-0.005)]$erf$[0.14(S_{\rm{int}}-1.0)]$ &\\
\hline
\end{tabular}
\end{table*}

A quantitative version of Figure~\ref{noisemap} is as shown in
Figure~\ref{noisehist} with three normalised
distributions of cube `noise' levels (characterised by $1$-percentile
rank).  The cube `noise' levels of Southern HIPASS are represented by the
solid line distribution;  the dashed and dotted line
represent cube `noise' level distributions of Northern HIPASS.  The
cube `noise' level of the
southernmost declination band in Northern HIPASS is shown by the dashed line
distribution and the cube `noise' level  of the rest of Northern HIPASS is shown by
the dotted line distribution.  The median `noise' level for the solid
line and dashed line distributions are at 42 mJy and 43 mJy
respectively and  the median `noise' level for the dotted line
distribution (northernmost HIPASS) is 55 mJy.  By this measure, the noise level has increased
by $31\%$.   

In conclusion, the `noisiness' observed in the Northern HIPASS cube
can be described quite accurately using the 1-percentile rank 
which measures the values of the pixel flux outliers in each cube.
As mentioned before, the increase in the 1-percentile rank values for
the northernmost declination band in Northern HIPASS is most likely 
a result of the increase in solar interference from increasingly lower elevation observations.

\section{Completeness and reliability of NHICAT}

The techniques used to calculate the completeness and reliability in
NHICAT are the same as the methods used for HICAT.  Detailed descriptions of the 
methods used to analyse the completeness and
reliability of HICAT can be found in  \citet{zwaan04}.

\subsection{Completeness of NHICAT}

Synthetic sources were inserted into all the Northern HIPASS cubes
 before  NHICAT was constructed  in
 order to measure the completeness of the resulting catalogue.  These
 synthetic sources were then extracted after the parameter finding process.
  In total, 774 non-extended synthetic sources were inserted into the
Northern HIPASS cubes.  These sources  represent a random sample of sources 
ranging in $50\%$ velocity width ($W_{\rm{50}}$) from 20 to 650 \kms\ , 
from 0.02 to 0.13 Jy in peak flux ($S_{\rm{p}}$) and from
300 to 10000 \kms\ in heliocentric velocity ($v_{\rm{hel}}$).  

The completeness of recovery can be easily estimated by measuring the
fraction of fake sources (recovered in each parameter bin), $D$:

\begin{equation}
D(S_{\rm{p}},W) = N^{\rm{fake}}_{\rm{recovered}}(S_{\rm{p}},W)/N^{\rm{fake}}(S_{\rm{p}},W)
\end{equation}

However, the completeness as a function of one parameter cannot be
effectively measured solely using $D$.  The completeness, $C$, of NHICAT
can be measured via the ratio of the number of {\em{detected}} real sources,
$N$, over  the {\em{true}} number of sources in each bin.  The {\em{true}} number of 
sources in each bin can be estimated by $N/D$. Since the number of sources 
in each bin differs and the parameter distribution of the fake sources may be different from the
 distribution of the real galaxies, this method cannot give a very good estimate 
of the completeness as a function of one parameter. A correction can be made by
 integrating over another parameter and applying a weighting to account for the different
number of sources in each bin.  As an example, $C(S_{\rm{p}})$
 can be estimated by integrating over $W$ :
\begin{equation}
C(S_{\rm{p}}) = \frac{\Sigma^{\infty}_{\rm{W=0}}N(S_{\rm{p}},W)}{\Sigma^{\infty}_{\rm{W=0}} N(S_{\rm{p}},W)/D(S_{\rm{p}},W)}
\end{equation}

Likewise, $C(W)$ and $C(S_{\rm{int}})$ have been calculated by integrating over $S_{\rm{p}}$ .
Figure~\ref{compl}  shows the
completeness of NHICAT as functions of $S_{\rm{p}}$, $W$ and $S_{\rm{int}}$ 
respectively where the solid lines are error function fits to the
datapoint. The error bars on the datapoints were determined using
bootstrap re-sampling and show 68 percent confidence levels.  The
error function fits and the completeness limits at $95\%$ confidence
levels are given in Table~\ref{comp}. Using the same method as \citet{zwaan04}, 
different fitting functions were tested in order to fit the completeness as a function of two 
parameters.  The completeness of NHICAT as a function of 
$S_{\rm{p}} $(Jy) and $S_{\rm{int}}$ (Jy \kms\ ) is also shown in Table~\ref{comp}.

The completeness ($C(S_{\rm{p}})$) limits at $95\%$
confidence level is at 68 mJy for HICAT, while NHICAT's completeness at
the same confidence level is 91 mJy.  It appears that to first order, the completeness 
of $S_{\rm{p}}$ scales with the noise level.   However, it would be too simplistic to assume
that the noise levels and source detection scale linearly .
It should also be noted that cosmic variance has not been taken into account.

\subsection{Reliability of NHICAT}

The reliability of NHICAT was measured by re-observing a subsample of
NHICAT sources in the narrowband mode at the Parkes Telescope.  As
with the reliability estimation of HICAT \citep{zwaan04}, a random
sample representing the full range of
NHICAT parameters was chosen, while giving preference to NHICAT detections with
low $S_{\rm{int}}$ and $S_{\rm{p}}$ (generally with $S_{\rm{int}} < 8$ Jy \kms\ and
$S_{\rm{p}} < 0.07$ Jy).  

\subsubsection{Narrowband observations}

In addition to calculating the reliability of NHICAT, the narrowband
observations were also used to remove spurious detections from the
catalogue.  As such, the less certain source detections (Flag=2) were observed as higher
priority.  In addition, definite source detections (Flag=1) were chosen randomly
by the observer for observation.   These narrowband
observations  took place over 4 observing sessions from July 2003 to February 2005. 

A spectral resolution of 1.65 \kms\ at $z=0$ was used by observing with
the narrow-band mode which consisted of 1024 channels over 8
MHz. Integration times were approximately 15 minutes for each source.  The
specific details of the observing mode used can be found in \citet{zwaan04}.
The narrowband observations were reduced using the {\sf AIPS++} packages {\sc
  livedata} and {\sc gridzilla} \citep{barnes01} as with the narrowband observations of HICAT.

The percentage of rejected sources in the NHICAT
narrowband observations was greater than for HICAT.  
 A likely explanation is that the
signal-to-noise level is less in the north and thus source detection algorithm is less
effective than in the south.
 
\begin{figure}
\begin{center}
\vspace{2pc}
\begin{tabular}{cc}
\includegraphics{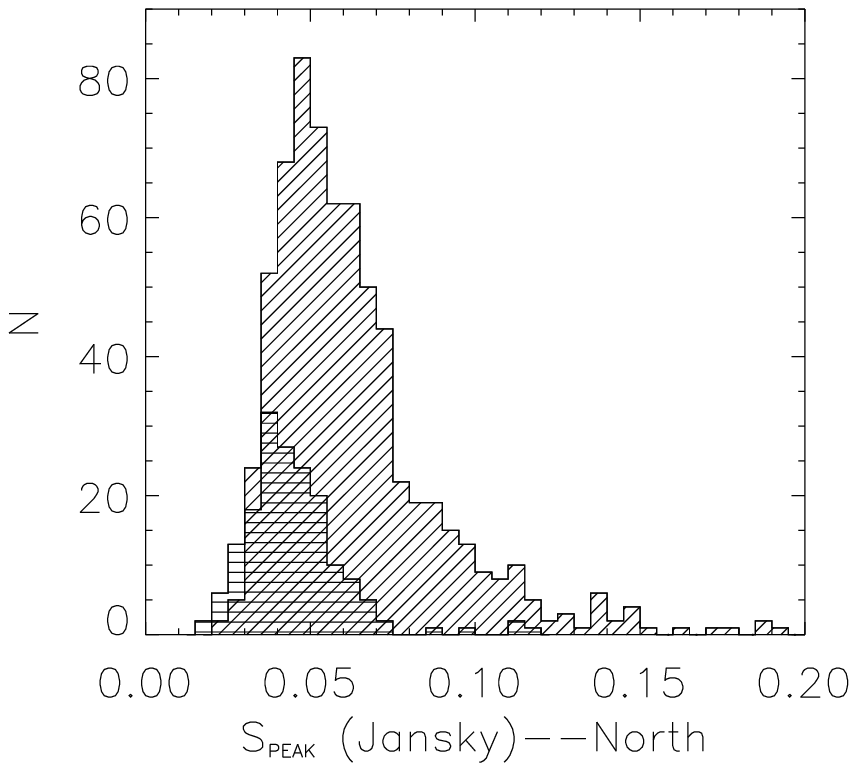} & \includegraphics{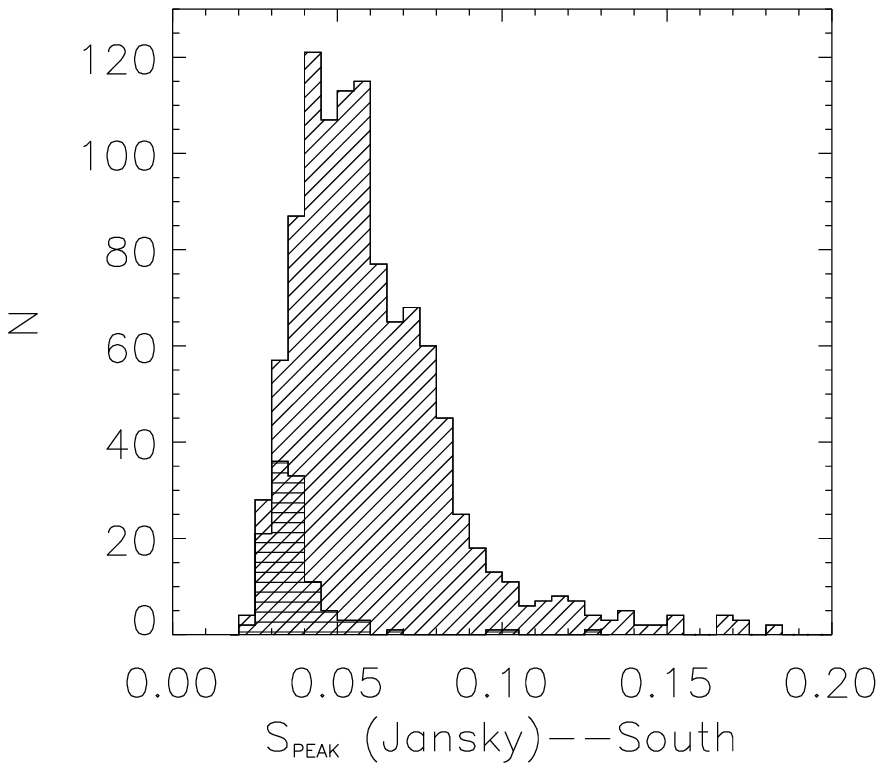} \\
\end{tabular}
\vspace{7pc}
\caption{Peak flux, $S_{\rm{peak}}$ (Jy), distributions of the follow-up observations in Northern HIPASS (left) and in Southern HIPASS (right).  The distributions marked with lines on a 45 degree angle represent the population of observed and confirmed detections, while, the distributions marked with horizontal lines represent the population of non-detected sources.}
\label{fupdist}
\end{center}
\end{figure}

A total of 857 sources were observed, of which 172 ($20\%$) 
were rejected, compared with the narrowband observations of HICAT where
less than $10\%$ of the observed sources were rejected.  Figure~\ref{fupdist}
 shows the peak flux distributions of the follow-up observations in both Northern 
and Southern HIPASS.  As can be seen from this figure, more sources were 
rejected at $S_{\rm{peak}} > 0.05$ Jy in the Northern follow-up observations than
in the Southern follow-up observations.  Also, the percentage of  sources with $S_{\rm{peak}} > 0.05$
observed in the Southern follow-up observations are higher than in the Northern
observations which may also explain the difference in the percentage rejected.

\subsubsection{Reliability measure}
As explained in \citet{zwaan04}, the reliability will improve when
more uncertain sources are removed after the narrowband follow-up
observations.  Therefore we start by examining the original
catalogue before unconfirmed sources were removed.  The ratio of
the number of confirmed sources ($N^{\rm{rel}}_{\rm{conf}}$) to the number of
observed sources ($N^{\rm{rel}}_{\rm{obs}}$) is defined to be :
\begin{equation}
T(S_p,W) = N^{\rm{rel}}_{\rm{conf}}(S_{\rm{p}},W) / N^{\rm{rel}}_{\rm{obs}} (S_{\rm{p}},W)
\end{equation}
The reliability $R$ as a function of a single parameter is the mean of
$T$, weighted by the number of sources in each bin.  For example the
reliability as a function of peak flux ($S_{\rm{p}}$) is:
\begin{equation}
R(S_{\rm{p}}) = \frac{\Sigma^{\infty}_{\rm{W=0}} N(S_{\rm{p}},W) \times
  T(S_{\rm{p}},W)}{\Sigma^{\infty}_{\rm{W=0}} N(S_{\rm{p}},W)}
\label{reli_eq}
\end{equation}
Similarly, $R(W)$ and $R(S_{\rm{int}})$ can be measured by integrating over
$S_{\rm{p}}$.  

\begin{figure*}
\begin{center}
\includegraphics{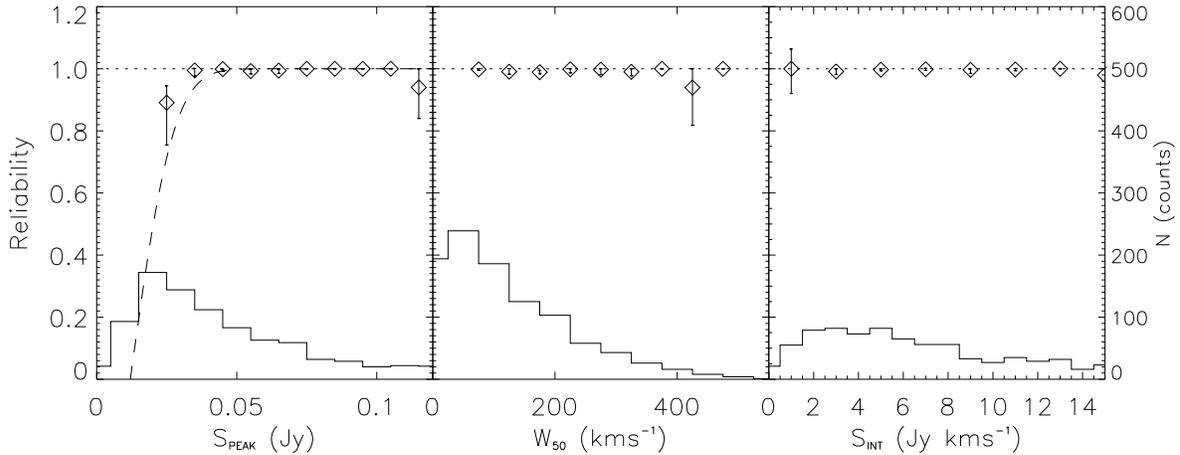}
\vspace{15pc}
\caption{Reliability of NHICAT as a function of $S_{\rm{p}}$, $W_{\rm{50}}$ and $S_{\rm{int}}$.  The histograms 
show the $S_{\rm{p}}$, $W_{\rm{50}}$ and $S_{\rm{int}}$ distributions of NHICAT sources.}
\label{reli}
\end{center}
\end{figure*}

\begin{figure*}
\begin{center}
\includegraphics{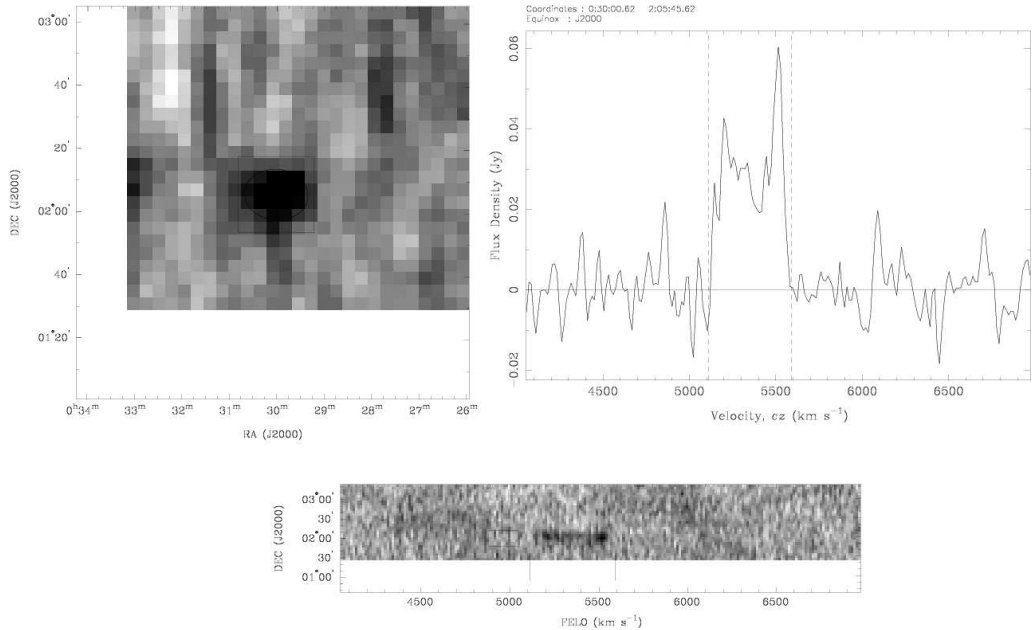}
\vspace{19pc}
\caption{Example of 3 data products available online at $\langle${\bf{\tt http://hipass.aus-vo.org}}$\rangle$ for source HIPASSJ0030+02. Clockwise from top left: Integrated intensity map, \HI\ spectra and a position-velocity projection intensity map.  }
\label{egmaps}
\end{center}
\end{figure*}

Since  $20\%$ of the observed sources in the initial NHICAT have been
rejected and removed from the catalogue, the reliability of the
catalogue has been improved by re-observing a subsample of the sources.

An estimate of the {\em{expected}} number of real sources has to be
made in order to calculate the reliability of NHICAT after the removal
of unconfirmed sources.  The expected number of sources is given by:
\begin{equation}
N_{\rm{expectreal}} = N_{\rm{confirmed}} + ( T \times N_{\rm{unobserved}})
\end{equation}
where $N_{\rm{unobserved}}$ is the number of sources that have not been
reobserved.

T can now be redefined as :
\begin{equation}
T(S_{\rm{p}},W) = \frac{N_{\rm{expectreal}}(S_{\rm{p}},W)}{N(S_{\rm{p}},W)}
\end{equation}
where $N(S_{\rm{p}},W)$ is the total number of sources in NHICAT excluding
the unconfirmed sources.  Equation~\ref{reli_eq} can now be used to
calculate the reliability of the final NHICAT.

Figure~\ref{reli} shows the $S_{\rm{p}}$, $W$ and $S_{\rm{int}}$ distributions of NHICAT sources
as well as the reliability of NHICAT as functions of $S_{\rm{p}}$, $W$ and $S_{\rm{int}}$ 
 where the dashed lines are error function fits to the
datapoint. It should be noted that 871 sources, 1002 sources and 764 sources are plotted in the 
$S_{\rm{p}}$, $W$ and $S_{\rm{int}}$ distributions, respectively. This is due to the fact that 
 the rest of the sources have $S_{\rm{p}}$ and $S_{\rm{int}}$ greater than the parameter ranges
given in Figure~\ref{reli}.  The error bars on the datapoints were determined using
bootstrap re-sampling and indicate 68 percent confidence levels.  The
error function fits and the completeness limits at $95\%$ confidence
levels are given in Table~\ref{reli}.

\begin{table}
\caption{Reliability of NHICAT.}
\label{reli}
\centering
\begin{tabular}{llc}
\hline
Parameter & Reliability Fit & C=0.95 \\
\hline
$S_{\rm{p}} $ (Jy) & erf$[58(S_{\rm{p}}-0.012)]$ & 0.036\\
\hline
\end{tabular}
\end{table}
It is interesting though to note that the $95\%$ level of reliability
in NHICAT is lower (in $S_{\rm{p}}$ and $S_{\rm{int}}$) than the $95\%$ level of reliability in HICAT. This result can be attributed to the fact that  a larger proportion of the original NHICAT have been  re-observed in the narrowband follow-up observations.


\section{Summary}

In the northern extension of HIPASS, which covers the region between
declinations $+2^{\circ} < \delta < +25^{\circ} 30\arcmin$, 1001 extragalactic
sources were found and catalogued in NHICAT. In addition an extra source  
found with $\delta$ slightly less than $+2^{\circ}$ (which was not detected in HICAT) has also been included into NHICAT.

NHICAT has been found to be $95\%$ complete at peak flux 95 mJy and at an
integrated flux 15 Jy \kms\ .  This catalogue is also reliable at
a $95\%$ level at peak flux 36 mJy.  The entire catalogue and source spectra 
will be made publicly-available online at  
$\langle${\bf{\tt http://hipass.aus-vo.org}}$\rangle$.  The catalogue parameters
presented in the online archive are the same 33 parameters detailed in Table 4 of
\citet{meyer04}.  An excerpt of the online catalogue is shown in Table~\ref{sample}.
The online archive also contains the detection spectra and integrated intensity
maps in various projections. Examples of these data products are shown in Figure~\ref{egmaps}.  In addition to the Aus-VO archive, the catalogue and spectra will also be submitted to the NASA/IPAC Extragalactic Database (NED).



\vspace{1cm}

\chapter{\bf{Acknowledgments.}}
The Multibeam system was funded by the Australia Telescope National
Facility (ATNF) and an Australian Research Council LIEF grant.  The HIPASS collaboration was partially supported by an ARC grant (DP0208618).  The
collaborating institutions are the Universities of Melbourne, Western
Sydney, Sydney and Cardiff, Research School of Astronomy and
Astrophysics at Australian National University, Jodrell Bank
Observatory and the ATNF.  The Multibeam receiver and correlator was
designed and built by the ATNF with assistance from the Australian
Commonwealth Scientific and Industrial Research Organisation Division
of Telecommunications and Industrial Physics.  The original low noise
amplifiers were provided by Jodrell Bank Observatory through a grant
from the UK Particle Physics and Astronomy Research Council.

We thank N. Bate, A. Karick, E. MacDonald, M. J. Pierce, R. M. Price,
N. Rughoonauth, S. Singh, D. Weldrake and M. Wolleben for their help with the Parkes narrow-band follow-up
observations.  Finally, we would like to thank the  staff at the
Parkes Observatory for all their support.

\bibliographystyle{mn2e}
\bibliography{mn-jour,paperef}

\footnotesize{
\begin{landscape}
{\setlength{\textwidth}{8.5in}
\begin{table*}
\begin{center}
\caption{Excerpt from NHICAT. Note that all the velocities are $cz$ and heliocentric. }
\label{sample}
\begin{tabular*}{50pc}{lllccccccccc}
\hline
Name & RA & Dec & $v_{50}^{\rm{max}}$ (\kms) &$v_{50}^{\rm{min}}$ (\kms)&$v_{20}^{\rm{max}}$ (\kms)&$v_{20}^{\rm{min}}$ (\kms)&$v_{\rm{mom}}$ (\kms)\\
&$v_{\rm{S_p}}$ (\kms)&$v_{\rm{gsr}}$ (\kms)&$v_{\rm{lg}}$ (\kms)& $v_{\rm{cmb}}$ (\kms)& $v_{\rm{lo}}$ (\kms) &$v_{\rm{hi}}$ (\kms)&$v_{\rm{speclo}}$ (\kms)\\
&$v_{\rm{spechi}}$ (\kms)&$v_{\rm{mask}}$ (\kms) &$W_{50}^{\rm{max}}$ (\kms)& $W_{50}^{\rm{min}}$ (\kms)&$W_{20}^{\rm{max}}$ (\kms)&$W_{20}^{\rm{min}}$ (\kms)& $S_{\rm{p}} (Jy)$ \\
&$S_{\rm{int}} (Jy \kms)$ &RMS (Jy) &RMS$_{\rm{Clip}}$ (Jy) &RMS$_{\rm{Cube}}$ (Jy)&Cube &Sigma (\kms)&Box size (\arcmin)\\
& Comment &Follow-up & Confused  &Extended &&& \\
\hline
HIPASSJ0030+02& 00:30:00.6 &02:05:46 	&	 762.0& 5509.0& 5347.5 &  765.3 	&		 5362.3\\
& 5514.5 &		 5370.3& 5344.9& 5336.2 & 5111.8& 5590.2& 4055.5\\
& 6973.1& 5112,5590& 			 361.8& 73.5& 434.8& 434.8 	&		 0.060\\
& 13.5 &		 0.0065& 0.0049& 0.01209 &			 389& 	 12& 7 \\
	&	 1 & 1 & 0&  0 &&&\\
HIPASSJ0033+02 & 00:33:44.3 & 02:40:37  &		 4340.8 & 4454.8 & 4390.4 &   4336.6 & 			 4402.7\\
 & 4478.3 	 &	 4411.8 & 4386.3 & 4377.6 &		 4242.8 & 4542.0 & 2989.5 \\
& 5941.5 & 4243,4542 & 			 225.9 & 94.6 & 248.0 & 248.0  &			 0.069\\
 & 9.5 &		 0.0080 & 0.0073 & 0.01222  &			 390 & 	 12 & 7 \\
& 		 1  & 1 &  0 &  0 &&&\\
HIPASSJ0142+02&	 01:42:28.4&	 02:56:20 &			 2988.9&	 1763.9&	 1764.0  &	 2989.1 &				 1762.8\\
&	 1744.6 &			 1772.0&	 1746.6 &	1737.9&			 1685.4&	 1843.3&	 321.3 \\
&	3221.2&	 1685,1843&	 			 80.6 &	80.6&	 109.9&	 109.9 &				 0.112\\
&	 9.0&		 0.0070&	 0.0058&	 0.01362 &				 392&	 	 12&	 7\\
 &	1&	  0&	  0 &	 0&	&&\\
HIPASSJ0150+02& 01:50:15.2& 02:18:58 &		 1503.8& 1695.8& 1696.6 &  1505.4 &			 1696.0\\
& 1677.9& 		 1703.0& 1677.6& 1668.9 &		 1618.9& 1750.0 &255.4\\
& 3127.1& 1619,1750& 			 73.0 &73.0& 94.9 &94.9 		&	 0.053\\
& 3.5 &	 0.0065& 0.0060& 0.01362 &			 392& 	 12& 7 \\
&		 1&  1 & 0&  0&&&\\
HIPASSJ0249+02& 02:49:06.4& 02:08:27 &		 1024.6& 1104.7& 1103.6 &  1023.3 &			 1104.5\\
& 1105.4 &		 1113.8& 1088.4& 1079.7& 			 1047.8& 1167.3 &-324.4 \\
&2536.0 &1048,1167 &			 58.6&58.6 &81.0& 81.0 &			 0.945\\
& 55.6 &	 0.2363 &0.0543 &0.01380 		&	 394 &	 12& 7 \\
	&	 1&  0&  0 & 0&&&\\
HIPASSJ0253+02& 02:53:48.6& 02:20:42 &		 3645.8& 6787.8& -9999.0 &  3634.9& 			 6730.6\\
& 6833.6 &		 6739.1& 6713.7 &6705.0&		 6532.9 &6924.2 &5129.9\\
 &8319.0& 6533,6924		&	 349.4& 235.8 &0.000& 0.000 &			 0.037\\
& 9.2 &		 0.0072 &0.0066 &0.01440& 			 395 	& 12& 7\\
 &		 1 & 1&  0&  0&&&\\
HIPASSJ0254+02& 02:54:05.6& 02:57:22 &		 6731.0& 2952.5& 3039.7  & -9999.0 &			 3036.2\\
& 2937.1& 		 3043.9& 3018.5& 3009.8&	 2864.7& 3212.5& 1490.0 \\
&4603.0 &2865,3212 &			 264.4& 93.4 &293.6 &293.6 		&	 0.090 \\
&17.1 	&	 0.0075& 0.0063& 0.01440 	&		 395 &	 12 &7\\
& 		 1 & 0&  0 & 0&&&\\
HIPASSJ0259+02& 02:59:48.2 &02:45:38 	&	 376.0& 2793.6 &2848.4 &  -9999.0 &			 2844.4 \\
&2762.3 &		 2850.4& 2825.0 &2816.4 &		 2668.9& 3025.0& 1516.0\\
& 4384.8& 2669,3025 &			 242.6& 137.5& 307.0 &307.0 &			 0.088\\
 &16.9 & 0.0075& 0.0069 &0.01440 &			 395 &	 12& 7 \\
&		 1&  0 & 0&  0&&&\\

\hline
\end{tabular*}
\end{center}
\end{table*}}
\end{landscape} }
\bsp

\label{lastpage}




%


\end{document}